\documentclass[12pt]{article}

\usepackage[colorlinks=false]{hyperref}
\usepackage[dvips]{graphicx}
\usepackage[usenames,dvipsnames]{color}
\usepackage{epsfig}
\usepackage{rotating}
\usepackage{amssymb}
\usepackage{amsmath,amsfonts}
\usepackage[latin1]{inputenc}
\usepackage{verbatim}
\usepackage[tableposition=top,font=small,labelfont=bf,format=hang]{caption}
\usepackage{booktabs}

\newtheorem{thm}{Theorem}

\newtheorem{prop}{Proposition}

\newcommand{\F}{\mathbb{F}}

\newenvironment{proof}{\noindent\textbf{Proof.}\quad}
{\hspace{\stretch{1}}%
\rule{1ex}{1ex}\\}

\begin{document}
 \author{ A. Bonnecaze\thanks{ alexis.bonnecaze@univ-amu.fr}, P. Sol\'e\thanks{ corresponding author, sole@enst.fr}, 
\thanks{ AB and  PS are with Aix Marseille Univ, CNRS, Centrale Marseille, I2M, Marseille, France. }}
\title{ The extended binary quadratic residue code of length $42$ holds a $3-$design}
\date{}
\maketitle

\begin{abstract}
 The codewords of weight $10$ of the $[42,21,10]$ extended binary quadratic residue code are shown to hold a design of parameters $3-(42,10,18).$ Its automorphism group is isomorphic to $PSL(2,41).$
 Its existence can be explained neither by a transitivity argument, nor by the Assmus-Mattson theorem.
\end{abstract}

\ \

{\bf AMS Math Sc. Cl. (2010):} 94 B15,  62K10

\section{Introduction}
The quadratic residue codes have been known to hold designs of large strength since the seminal paper \cite{AM}, which constructs in particular the Witt designs by coding theoretic techniques.
The argument in that famous paper (see also \cite{AMT} for background)
is the so-called Assmus-Mattson theorem
which derives the existence of designs from certain facts on the weight distribution of the dual code \cite[Chap. 6, Th. 29]{MS}.

In the present note we produce, by electronic calculation, a $3-(42,10,18)$ design which is not in the table \cite[pp. 82--83]{CD}, cannot be derived from the Assmus-Mattson Theorem, and does not follow by the standard
transitivity argument. Its blocks are the supports of the codewords of weight $10$ in  the $[42,21,10]$ extended binary quadratic residue code. Its automorphism group coincides with that of that code.
Thus it is isomorphic to $PSL(2,41)$ which is not $3$-homogeneous. Note that $PSL(2,q)$ is  $3$-homogeneous in general for $q\equiv 3 \pmod{4}$ \cite{D}. Thus our design cannot be obtained by the methods of 
\cite{C+}. Neither can it be obtained from the orbits glueing methods of \cite[Th. 5]{BR}, or \cite[Th. 5.1]{LTW}, since $41 \equiv 9 \pmod{16}.$ Even though the code we
are using is a formally self-dual even code, the results of \cite{KP} do not apply, because the minimum distance is not high enough.
\section{Background material}
\subsection{Permutation groups}
A permutation group $G$ acting on a set $X$ is  {\em transitive} on $X$ if it has a single orbit on $X.$ Given an integer $s$ it is {\em  $s$-fold transitive} if it is transitive on ordered $s$-uples of 
distinct elements of $X.$ It is {\em  $s$-fold homogeneous} if it is transitive on unordered $s$-uples of distinct elements of $X.$
Let $p$ denote an odd prime. The group $PSL(2,p)$ is defined in its action on the projective line $\F_p \cup \infty$ as the set of all transforms $$y\mapsto \frac{a+by}{c+dy}$$
with $ad-bc=1,$ and the conventions $\frac{0}{0}=0$ and $\frac{\neq 0}{0}=\infty.$ See \cite[Chap. 16, \S 5]{MS}, or \cite[Chap 17]{PH} for background.
\subsection{Designs}
 A block design of parameters $t-(b,v,k,r, \lambda)$ ( shortly a $t-(v,k, \lambda)$ design) is 
an incidence structure $(\mathcal{P},\mathcal{B},I)$ satisfying the following axioms.

\begin{enumerate}
\item The set $\mathcal{P}$ has $v$ points.
\item The set $\mathcal{B}$ has $b$ blocks.
 \item Each block is incident to $k$ points.
 \item Each point is incident to $r$ blocks.
 \item Each $t$-tuple of points is incident in common with $\lambda$ blocks.
\end{enumerate}

\subsection{Codes}
Let $\F_2=\{0,1\}$ denote the finite field of order $2.$
A binary code of length $n$ is a $\F_2$ subspace of $\F_2^n.$
The {\em weight} of a vector of $\F_2^n$ is the number of its nonzero coordinates.
The {\em weight distribution} of a code $C$ is the sequence $A_w$ of number of codewords of $C$ of weight $w.$
It is written in Magma \cite{M} notation as the list with generic element $\langle w,A_w \rangle$ where $w$ ranges over the weights of $C.$
A binary code is {\em cyclic} if it is invariant under the cyclic shift. Cyclic codes are in one to one correspondence with ideals of 
the residue class ring $\F_2[x]/(x^n-1).$ The generator polynomial of a cyclic code is then the generator of the corresponding ideal. The
{\em Quadratic residue codes} are the cyclic codes of length $p,$ with $p$ an odd prime, defined for $p\equiv \pm 1 \pmod{8}$ by the generator polynomial
of degree $\frac{p-1}{2}$
$$ \prod_{r=\Box}(x-\alpha^r),$$ with $\alpha$ a primitive root of order $p.$ 
Since $2$ is a quadratic residue modulo an odd prime $p$ this polynomial is indeed in $\F_2[x].$ As a simple example consider 
the case of $p=7$ when the polynomial $$(x-\alpha)(x-\alpha^2)(x-\alpha^4)=x^3+x+1$$ generates the Hamming code $[7,4,3].$
See \cite[Chap.16]{MS} for background.

The {\em automorphism} group a binary code of length $n$ is the subgroup of the symmetric group on the $n$ coordinate places that leaves the code wholly invariant.

The Assmus-Mattson Theorem is as follows for binary codes \cite[Chap. 6, Th. 29]{MS} .

{\thm (Assmus-Mattson) If $C$ is a binary $[n,k,d]$ code such that the weight distribution of its dual code contains at most $d-t$ nonzero weights $\le n-t$
then the codewords of  weight $d$ of $C$ form a $t$-design.}

A folklore theorem is that if the automorphism group of a binary code is $t$-homogeneous then the codewords of any given weight $\ge t$ hold $t$-designs \cite[p. 308]{PH}.
.
\section{Construction}
Let $Q$ be the binary  quadratic residue code of prime length $41,$ and denote by $C$ its extension by an overall parity-check. The code $C$ is not self-dual, but it is equivalent to its dual
\cite[Chap. 16, Fig. 16.3]{MS}. Designs in even formally self-dual codes have been explored in \cite{KP}, but the codes there are extremal, which would require a $[42,21,12].$

The weight distribution of  $C$ is also its dual weight distribution 
$$[ \langle0, 1\rangle, \langle10, 1722\rangle, \langle12, 10619\rangle, \langle14, 49815\rangle, \langle16, 157563\rangle, \langle18, 341530\rangle,$$ $$ \langle20,
487326\rangle, \langle22, 487326\rangle, \langle24, 341530\rangle, \langle26, 157563\rangle, \langle28, 49815\rangle, \langle30, 10619\rangle,$$ $$
\langle32, 1722\rangle, \langle42, 1\rangle ].
$$
Thus the Assmus-Mattson theorem, which requires at most $10-3=7$ non zero dual weights $\le 42-3=39,$ cannot apply here.

{\thm The $1722$ codewords of weight $10$ of $C$ hold a $3-(1722,42,10, 410, 18)$ block design $\mathcal{D}.$} \\ \ \

\begin{proof}
 Magma computation \cite{M}.
\end{proof}

Its automorphism group is a permutation group of order $34440$ with generators
$$(3, 30, 29, 31)(4, 9, 18, 7)(5, 24, 25, 17)(6, 22, 38, 42)(8, 34, 11,28)$$ $$(10, 36, 16, 33)(12, 32, 23, 21)(13, 15, 14, 41)(19, 20, 39, 27)(26,
        35, 40, 37),$$
 $$   (3, 8, 6, 33, 15, 29, 11, 38, 36, 41)(4, 35, 27, 21, 5, 18, 37, 20, 32,
        25)$$ $$(7, 26, 39, 23, 17, 9, 40, 19, 12, 24)(10, 14, 31, 28, 42, 16, 13,
        30, 34, 22),$$
  $$  (1, 32)(2, 21)(3, 36)(5, 10)(6, 26)(7, 38)(8, 20)(9, 22)(11, 25)(12, 23)(13,
        33)(14, 28)$$ $$(15, 39)(16, 27)(17, 30)(18, 37)(19, 31)(24, 41)(29, 34)(40,
        42),$$
   $$ (2, 32, 23, 12, 36, 29, 37, 14, 24, 10, 8, 15, 40, 6, 4, 31, 41, 18, 26,
        19)$$ $$(3, 42, 11, 34, 9, 28, 20, 17, 30, 33, 39, 7, 16, 22, 25, 38, 35, 27,
        5, 21),$$ 
        
        This group can be characterized as an abstract group as follows.
    {\prop   The automorphism group $G$ of $\mathcal{D}$ is isomorphic to $PSL(2,41).$}\\ \ \
    
    \begin{proof}
     The automorphism group of $C$ is known to be isomorphic to $PSL(2,41)$ by \cite[Chap. 17]{PH}. Since the blocks of $\mathcal{D}$ consist of all the codewords of weight $10$ of $C$, they are wholly invariant under 
     the automorphism group of $C.$ A Magma computation \cite{M} shows that the order of the automorphism group  $\mathcal{D}$  is $34440=\frac{1}{2}41\times (41^2-1),$ the known order of $PSL(2,41)$ 
     \cite[Chap. 16, Th. 9 (b)]{MS}. The result follows.
    \end{proof}

    The group $G$ is {\em not} $3$-homogeneous.
    
      {\prop The group $G$ partitions the set of triples into two orbits of equal size. }\\ \ \
      
    \begin{proof}
     A Magma computation shows that the orbits of $\{1,2,3\},$ and $\{1,3,8\},$ are disjoint and both of size $5740={ 42 \choose 3}/2.$
    \end{proof}

{\bf Remarks:}  
     \begin{enumerate}
      \item  The  codewords of weights $9$ and $10$ in the quadratic residue code of length $41$ hold $2$-designs of parameters $2-(41,9,18)$ and $2-(41,10,72)$ 
      They cannot be explained by the Assmus-Mattson theorem or group action. The $2-(41,9,18)$ is a derived design of $\mathcal{D}.$
      \item It can be shown that the linear span of $\mathcal{D}$ is $C.$ This gives an alternative proof of Proposition 1.
      \item The codewords of minimum weight in the extended quadratic residue code of length $74$ do not hold 3-designs.
      \item The codewords of weight 10 and 12 in the duadic codes of length $74$ of \cite{LMP} do not hold 3-designs.
     \end{enumerate}

\section{Conclusion}
In this work, we have constructed a new $3$-design in the minimum weight codewords of a quadratic residue codes. 
Its existence cannot be explained by the Assmus-Mattson Theorem, nor by group action.

The arguments are based on machine calculation. It would be more satisfying to have a conceptual 
derivation. A method by multivariate weight enumerators and invariant theory in the spirit of \cite{BMS}, would be more conceptual, but certainly not computer-free.
The fact that the low weight codewords in the extended quadratic residue code of length $74$ or in the duadic codes of that length do not hold 3-designs suggests
that the design in this note is  exceptional enough that a general conjecture is impossible to formulate.

{\bf Acknowledgement:} The authors thank Professor Jenny Key for helpful discussions.

\end{document}